\documentclass[aps,preprint,groupedaddress,prl,longbibliography]{revtex4-1}%
\usepackage{amssymb}
\usepackage{amsfonts}
\usepackage{amsmath}
\usepackage{graphicx}%
\usepackage{xcolor}
\begin{document}

\title{Pulse strategy for suppressing spreading on networks}
\author{Qiang Liu}
\email{Q.L.Liu@tudelft.nl}
\author{Xiaoyu Zhou}
\author{Piet Van Mieghem}
\email{P.F.A.VanMieghem@TUDelft.nl}
\affiliation{Faculty of Electrical Engineering, Mathematics and Computer Science, Delft University of Technology, Delft, the Netherlands}
\begin{abstract}
In networked spreading models, each node can infect its neighbors and cure spontaneously. The curing is assumed to occur uniformly over time. A pulse immunization/curing strategy is more efficient and broadly applied to suppressing spreading processes. We model the epidemic process by the basic Susceptible-Infected (SI) process with a pulse curing and incorporate the underlying contact network. The mean-field epidemic threshold of the pulse SI model is shown to be $\frac{1}{\lambda_1}\ln\frac{1}{1-p}$, where $\lambda_1$ and $p$ are the largest eigenvalue of the adjacency matrix and the fraction of nodes covered by each curing, respectively. Compared to the extensively studied uniform curing process,  we show that the pulse curing strategy saves about $36.8$\%, i.e. $p\approx 0.632$, of the number of curing operations invariant to the network structure. Our results may help related policy makers to estimate the cost of controlling spreading processes.
\end{abstract}
\maketitle
\section{Background}
Viral spreading processes cause enormous losses of life. During the period 2009.04 to 2010.08, $18500$ laboratory-confirmed deaths are reported, while $284500$ deaths are estimated due to the pandemic influenza A H1N1 \cite{DAWOOD2012687}. Cyber-criminals earned around \$$100$ million per year by spreading an exploit kit, \emph{Angler}, in computer systems \cite{Cisco}. A recent study shows that false news spreads faster and more broadly than true news online \cite{Vosoughi1146}. The suppression of spreading processes is thus necessary for many circumstances, but consumes resources, e.g. budget in disease control or computational resources in detecting computer virus. Based on the data from the World Health Organization, around $19.9$ million children under the age of one still cannot receive the basic diphtheria-tetanus-pertussis (DTP3) vaccine and the coverage level of DTP3 for infants is only about $85$\% in 2017. Cisco reported \cite{Cisco} that $83$\% of the Internet of Things devices are not patched to be immunized against cyber attacks.

The trade-off between the cost and performance introduces a strategy design problem in suppressing spreading. The pulse strategy was first proposed to control the epidemic of measles \cite{Dabbagh2018} by periodically and synchronously vaccinating several age cohorts instead of uniformly vaccinating each individual at certain ages \cite{Agur11698,STONE2000207}. In 1995, India introduced the National Immunization Days, which is a pulse strategy, to control the spread of polio~\cite{India}. Comparing to the straightforward uniform strategy, the pulse strategy shows a better performance \cite{NOKES199714}.

On the other hand, the spreading process is also a focal topic in network science because the underlying contact network influences the spreading process non-trivially. For example, the epidemic threshold, which is determined by the network structure, of scale-free networks converges to zero with the network size under the mean-field approximation \cite{PhysRevLett.86.3200,10.2307/27795079,Chung2003,van_mieghem_virus_2009,PhysRevLett.105.218701}. However, the spreading processes studied on networks are generally Markovian, which means that the infection and curing occur both uniformly over time \cite{van_mieghem_2014}. Since networked systems exist broadly, it is necessary to study the pulse strategy on networks. From a point of view of network models, the pulse strategy reduces the reinfections between neighbor nodes. If the curing occurs for all nodes at the same time, then no reinfection happens and the spreading process is immediately killed forever. Although the curing may not cover the whole population, synchronous curing introduced by the pulse strategy still eliminates a substantial part of reinfections between neighbors and thus leads to better performance compared to a uniform, asynchronous curing strategy. Thus, one may wonder how to quantify the effectiveness of the pulse strategy. The most reasonable way is evaluating the reduction of the number of curing operations by using the uniform strategy as a benchmark. In the following, we consider the most basic spreading model on networks: the Susceptible-Infected (SI) process and evaluate the pulse strategy performed on the SI model.
%Pulse curing or vaccination is a common strategy to suppress epidemics by disease control agencies.  If the curing operation is perfect and synchronously covers the whole population, the disease will be immediately eliminated. However, each curing action only covers a fraction of the whole population or the curing effect is not perfect due to a limited effectiveness, which gives the spreading process the opportunity of surviving just as the asynchronous curing, e.g. Poisson curing process. In this paper, we study the effect of the pulse curing and indicate the relationship between the limitation of the curing resources and the epidemic threshold.
\section{The Model}
In the networked spreading process, each node in the network is either infected or susceptible (healthy). Each infected node can infect each healthy neighbor by a Poisson process with rate $\beta$. We assume that each node is cured with rate $\delta$. Thus, for the pulse curing strategy, the curing happens every $1/\delta$ time units, i.e. the nodes can only be cured at time $k/\delta$ for $k=1,2,\ldots$. The curing has a successful probability $p$ turning an infected node into healthy. Equivalently, one can think that each node can be cured certainly but only a fraction $p$ of nodes are randomly chosen to be cured. We define the effective infection rate $\tau\triangleq \beta/\delta$.

The difference between the above pulse curing SI model and the extensively studied Markovian SIS process~\cite{pastor2015epidemic} is: Each node in the SIS model is asynchronously cured by a Poisson process with rate $\delta$ and $p=1$, which represents a uniform curing strategy. In the SIS process on networks, there exists an epidemic threshold~\cite{van_mieghem_virus_2009,PhysRevLett.105.218701} under the $N$-Intertwined mean-field approximation $\tau_c^{(1)}=\frac{1}{\lambda_1}$ where $\lambda_1$ is the largest eigenvalue of the adjacency matrix of the network. If $\tau>\tau_c^{(1)}$, then the process is in an endemic phase in the steady state but if $\tau<\tau_c^{(1)}$, then the process converges to the all-healthy state. In the pulse curing strategy, the coverage $p<1$, because if $p=1$ then synchronous curing kills the spreading immediately. The average numbers of curing operations in the uniform Poisson curing and the pulse curing are $\delta$ and $\delta p$, respectively, for each node during one unit of time. In the following, we analyze the pulse curing effect on epidemic processes on networks under the mean-field theory to derive the epidemic threshold. Our main finding is that when $p=1-1/\mathrm{e}\approx  0.632$, the pulse curing is equivalent to a Poisson curing process with the same curing rate $\delta$.
\subsection{Mathematical analysis}
We represent the time $t$ in the form of $t=k/\delta+t^*$, where $t^*\in[0,1/\delta)$. For $t^*\neq 0$, only infection happens and the mean-field equation of node $i$ is
\begin{equation}\label{eq_meanfieldSI}
  \frac{dv_i(k/\delta+t^*)}{dt^*}=\beta\left[1-v_i(k/\delta+t^*)\right]\sum_{j=1}^{N}a_{ij}v_j(k/\delta+t^*)
\end{equation}
where $v_i(k/\delta+t^*)$ is the probability that node $i$ is infected at time $t=k/\delta+t^*$ and $a_{ij}\in \{0,1\}$ is the element of adjacency matrix of the size-$N$ network. The probability $v_i(k/\delta+t^*)$ is discontinuous at $t^*=0$ for all $k$ where curing happens: $\lim\limits_{t^*\rightarrow 0}v_i(k/\delta+t^*)=v_i(k/\delta)$ and $\lim\limits_{t^*\rightarrow 1/\delta}v_i(k/\delta+t^*)\neq v_i((k+1)/\delta)$. Equation (\ref{eq_meanfieldSI}) is under the mean-field method because we omit the correlation of the infection state between neighbors just as in the Markovian SIS process~\cite{PhysRevE.89.052802}. Since the curing probability of each node at $k/\delta$ is $p$, the pulse curing process is governed by the following equation,
\begin{equation}\label{eq_curing_periodic}
  v_i\left(\frac{k+1}{\delta}\right)=(1-p)\lim\limits_{t^*\rightarrow 1/\delta}v_i\left(\frac{k}{\delta}+t^*\right)
\end{equation}

In our previous work~\cite{PhysRevE.97.022309}, we introduced the bursty SIS model where the infection happens periodically with rate $\beta$ and the curing is a Poisson process. In the bursty SIS model, the relationship between the infection probability of each node at the start $t^*=0$ and the end $t^*\rightarrow 1/\beta$ of the same time interval is explicitly known as an exponentially decreasing function. In pulse curing, the relationship between $v_i(k/\delta)$ and $\lim\limits_{t^*\rightarrow 1/\delta}v_i(k/\delta+t^*)$ is described by (\ref{eq_meanfieldSI}) which does not have an explicit solution for general networks~\footnote{Only for the regular graph when the initial condition of each node is identical, there is a explicit solution for (\ref{eq_meanfieldSI}). One may verify that the results in the $d$-regular graph is $v_i(k/\delta)=(1-p)-1/\mathrm{e}^{d\tau}$. Let $v_i(k/\delta)=0$ and the threshold is $1/d\ln\frac{1}{1-p}$ which is consistent with (\ref{eq_threshold})}. However, since we only care about the regime where $v_i(k/\delta+t^*)\rightarrow 0$ to derive the epidemic threshold, we can first linearize Eq.~(\ref{eq_meanfieldSI}) around $v_i(k/\delta+t^*)=0$ for all $i$ and obtain
\begin{equation}\label{eq_linerizeSI}
  \frac{d\mathbf{v}(k/\delta+t^*)}{dt^*}=\beta A \mathbf{v}(k/\delta+t^*)
\end{equation}
where the infection probability vector $\mathbf{v}(k/\delta+t^*)\triangleq [v_1(k/\delta+t^*),\ldots,v_N(k/\delta+t^*)]^T$. The general solution~\cite[p.~209]{van_mieghem_2014} of (\ref{eq_linerizeSI}) is $\mathbf{v}(k/\delta+t^*)=\mathrm{e}^{\beta A t^*}C$ where $C=\mathbf{v}(k/\delta)$ is the initial value vector at $t^*=0$. Thus, the solution of Eq.~(\ref{eq_linerizeSI}) evaluated at $t^*\rightarrow 1/\delta$ is
\begin{equation}\label{eq_linerizeSIsolutionatEnd}
  \lim\limits_{t^*\rightarrow 1/\delta}\mathbf{v}(k/\delta+t^*)=\mathrm{e}^{\tau A}\mathbf{v}(k/\delta)
\end{equation}
Substitute (\ref{eq_linerizeSIsolutionatEnd}) into the curing equation (\ref{eq_curing_periodic}), we obtain
\begin{equation}\label{eq_discrete}
  \mathbf{v}\left(\frac{k+1}{\delta}\right)=(1-p)\mathrm{e}^{\tau A}\mathbf{v}\left(\frac{k}{\delta}\right)
\end{equation}
When the largest eigenvalue of $(1-p)\mathrm{e}^{\tau A}$, which is $(1-p)e^{\tau \lambda_1}$, is smaller than $1$, the infection probability $\mathbf{v}\left(\frac{k}{\delta}\right)$ converges to zero in the long run due to (\ref{eq_discrete}). Thus, let $(1-p)e^{\tau \lambda_1}=1$ and we obtain the epidemic threshold
\begin{equation}\label{eq_threshold}
  \tau_c^{(p)}\triangleq\frac{1}{\lambda_1}\ln\frac{1}{1-p}
\end{equation}
If $\tau>\tau_c^{(p)}$, then the spreading can persist on the network. If $\tau<\tau_c^{(p)}$, then the spreading disappears in the long run.

The Markovian SIS process with a uniform Poisson curing process has a mean-field epidemic threshold $\frac{1}{\lambda_1}$. When $\ln\frac{1}{1-p}=1$, i.e. $p=1-1/\mathrm{e}\approx 0.632$, the pulse curing is equivalent to the Poisson curing process in the traditional SIS model on any networks. Thus, to eliminate the spreading, the pulse strategy only consumes $63.2$\% of the number of curing operations of the uniform strategy, since the curing rates $\delta$ of the two strategies are equal. In the next section, a typical example shows that even above the epidemic threshold, the two strategies are comparable, if $p=0.632$.
\subsection{Simulation: above the epidemic threshold}
\begin{figure}
    \centering
    \includegraphics[width=1\textwidth]{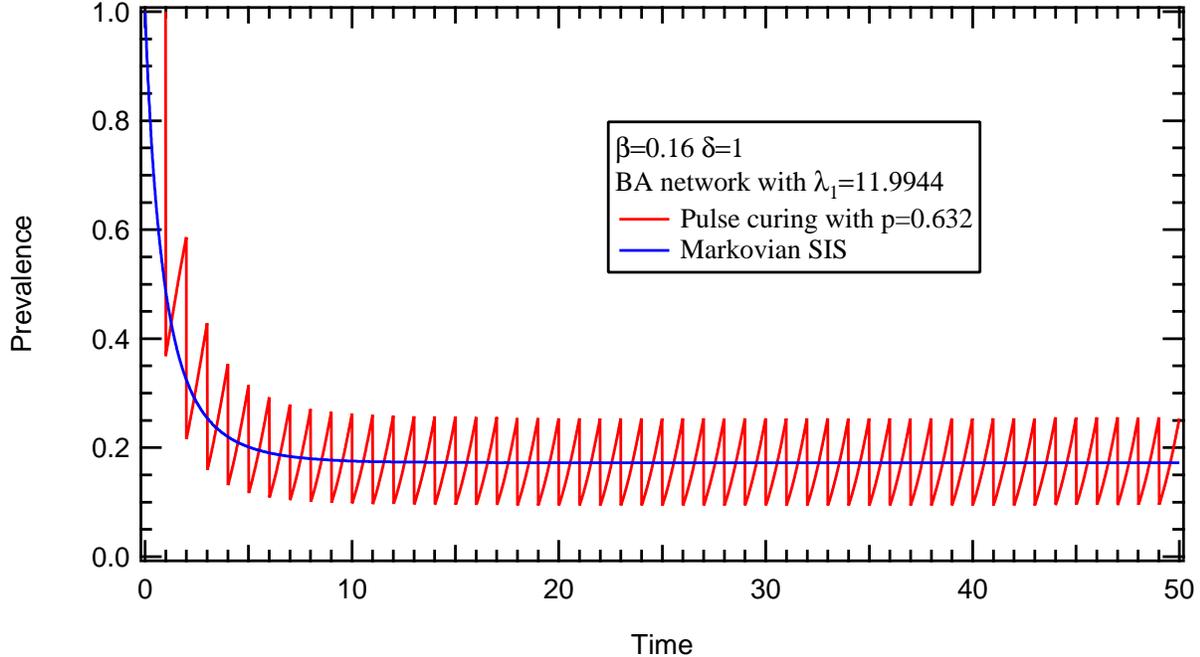}
    \caption{The prevalence of the Markovian SIS model and the pulse curing model obtained by averaging $10^5$ simulated realizations. The simulation is performed on a $500$-node network generated by the Barab\'asi-Albert model. The curing probability is set to be $p=0.632$ for the pulse strategy.}\label{prevalence}
\end{figure}

In Fig.~\ref{prevalence}, we show the prevalence of the Markovian SIS model and the pulse curing model with $p=0.632$, both with $\beta=0.16$ and $\delta=1$ on a Barab\'asi-Albert network \cite{Barabsi509}. The effective infection rate $\tau=0.16$ is above the epidemic threshold $1/\lambda_1=0.0834$. The prevalence, which is the average fraction of the infected nodes, of the Markovian SIS model is exactly centered at the middle of the prevalence generated by the pulse curing SI model. Figure~\ref{prevalence} indicates that the two curing processes are equivalent to some extents at $p=0.632$ even above the epidemic threshold.
\section{Conclusion}
We quantified the effect of the pulse strategy for suppressing spreading processes on networks. We show that the pulse strategy consumes $63.2$\% of the total number of curing operations, required by the uniform strategy to achieve an equivalent effect. This reduction of cost is invariant to the underlying network structure in the mean-field approximation. Our results may help the related agencies, e.g. disease control centers or computer security teams, to make policies or estimate resources budget in their tasks.
\bibliography{bibitem}
\end{document}